# Multiphysics model of chemical aging in frictional contacts

*Zhuohan Li and Izabela Szlufarska*

Department of Materials Science and Engineering, University of Wisconsin-Madison, Madison, 53706-1595, USA

**Abstract**
An increase of static friction during stationary contacts of two solids due to interfacial chemical bonding has been reported in multiple experiments. However, the physics underlying such frictional aging is still not fully understood because it involves multiple physical and chemical effects coupled with each other, making direct interpretation of experimental results difficult. Here, we develop a multiphysics chemical aging model that combines contact mechanics, mechanochemistry, and interfacial chemical reaction kinetics. Our model predicts that aging is proportional to normal loads in a low-load regime and becomes nonlinear at higher loads. We also discovered a nonmonotonic temperature dependence of aging with a peak near room temperature. In addition, our simulations provide insights into contributions from specific physical/chemical effects on the overall aging. Our model shows quantitative agreement with available single-asperity experiments on silica-silica interfaces, and it provides a framework for building a chemical aging model for other material systems with arbitrary types of physical and chemical effects involved.

**Main text**

Solid-solid frictional interfaces can undergo significant evolution over the time they are held in a stationary contact prior to sliding. This so-called frictional aging [1–5] is known to play a critical role in nucleation and recurrence of earthquakes [5], and also has a large influence on the performance and durability of microelectromechanical systems [6–8]. In general, aging has been attributed either to a change in contact area due to plastic deformation and/or to the change in quality of the interface due to chemical strengthening of the interface. In this study we focus on the role of chemical aging in friction. Possible mechanisms behind this phenomenon discussed previously in literature include formation of covalent bonds [9] and capillary condensation [10]. Chemical aging in friction was isolated for the first time by Li *et al.* [9] in atomic force microscopy (AFM) experiments. In this work, the authors reported a logarithmic increase of static friction with the hold time between an amorphous silica tip and an amorphous silica substrate. The underlying mechanism was later revealed by a theoretical study [11] which showed that formation of siloxane bonds across the hydroxylated silica-silica interface [12] alone can lead to the logarithmic aging based on the following reaction

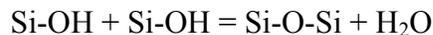

Si-OH + Si-OH = Si-O-Si + H$_2$O

Recently, AFM experiments by Tian *et al* [13] revealed that the amount of frictional aging increases linearly with the applied normal load. This linear dependence was attributed to the contact mechanics effect, i.e., to an almost linear relationship between the contact area and the normal load at low loads. This explanation is plausible, however, if contact mechanics truly plays an important role in aging, there should be a non-linear dependence of aging on normal load at high loads, which effect was not observed within the range of normal loads reported in Ref. [13].

In addition to the effect of normal load, temperature should also have a major influence on the chemical aging, since both formation and breaking of interfacial chemical bonds are thermally activated processes. Temperature dependence of static friction with chemical aging has not yet been reported. On the other hand, thermal effects have been widely studied in the context of kinetic stick-slip friction [14–16] and in many cases stick-slip friction can be successfully described by a traditional Prandtl-Tomlinson (PT) model [17,18]. However, PT model does not take into account the evolution of interfaces during stick phase and it fails to describe the non-monotonic temperature dependence of kinetic friction, which was sometimes observed in experiments [19,20]. Two different numerical models were later developed to address these issues and these models considered evolution of contact states. The so-called "mechano-kinetic model" [19,20] assumes that there are two competing processes of formation and breaking of "microscopic contacts" across the interface during the stick phase. The second model extends the traditional PT theory by assuming two stick states, i.e., a strongly bound state and a weakly bound state, with different energy barriers to the slip events [21,22]. Both numerical models are able to reproduce experimental results reasonably well. However, the physical origins of the "microscopic contacts" in the former model or "two stick states" in the latter model were not defined [23].

Here, in order to unravel the multi-physics nature of chemical aging, we build on our kinetic Monte Carlo (kMC) simulations of chemical aging [11] to develop a new multi-physics chemical aging model, which combines contact mechanics [24], mechanochemistry [11,25], and interfacial chemical reaction kinetics [11]. We choose a single asperity silica-silica interface as a model materials system in order to validate the predictions against available AFM experiments [13] and because silica contacts are relevant in multiple fields of science, such as geology [5] and semiconductor wafer bonding [26].

In order to account for contact mechanics effects (i.e., the dependence of contact area $A$ on normal load $F_N$) we assume a single asperity contact that is adhesive and elastic [9,13]. In the case of silica, such contact is well described by Derjaguin, Muller and Toporov (DMT) theory [24] [13]

$$A = \pi a^2 = \pi \left(\frac{3R}{4E^*}(F_N + F_{adh})\right)^{2/3} \qquad (1)$$

where $a$ is the contact radius, $R$ is the AFM tip radius, and $F_{adh}$ is the pull-off force. $E^* = [2(1-v^2)/E]^{-1}$ is the effective Young's modulus, where $E$ and $v$ are Young's modulus and the Poisson ratio, respectively. The interface is divided into a square grid with element size of $1/\rho_{OH} \approx 0.204$ nm$^2$, where $\rho_{OH} = 4.9$ OH/nm$^2$ is the density of OH group on fully hydroxylated silica surface under standard temperature and pressure conditions [27].

In the kMC model, we also include several mechanochemical effects on the interfacial chemical reaction that take place during stationary contact. First, we assume the energy barrier to bond formation varies across the interface [11] and depends on the local contact pressure following the Eyring relationship [11]. The local energy barrier to bond formation $E_{b,form}(x,y)$ in the presence of local pressure $P(x,y)$ is related to the local intrinsic energy barrier to bond formation $E_{b,form,i}(x,y)$ (i.e., the energy barrier in the absence of pressure) through the following relation

$$E_{b,form}(x,y) = E_{b,form,i}(x,y) - \Delta V P(x,y) \qquad (2)$$

where $\Delta V$ is the activation volume. From DMT model, the local contact pressure can be expressed as

$$P(x,y) = \frac{3(F_N + F_{adh})}{2\pi a^2}\sqrt{1-(r/a)^2} \qquad (3)$$

where $r = \sqrt{x^2 + y^2}$ is the in-plane distance from the tip center. Applicability of Eyring's theory to reactions on a silica-silica surface was previously verified using density functional theory (DFT) calculations [11]. It should be noted that the Eyring relationship is sometimes used as a mean effect under apparent normal pressure ($\bar{P} = F_N/A_0$)) [13,28,29], whereas in our model we explicitly consider the effect from local pressure on each reaction site.

Another mechanochemical effect that needs to be considered in the model is the elastic interaction between neighboring reaction sites/bonds. Previous DFT calculations [11] revealed that the formation of a siloxane bond most of the time increases the energy barrier to bond formation at neighboring reaction sites due to the elastic interaction mediated by deformation of tetrahedra in the bulk silica structure. Here, we assume that the change in the reaction energy barrier $\Delta E_{b,form}$ due to elastic interactions between neighboring reaction sites follows a uniform distribution determined based on the same expression as used in Ref. [11]

$$\Delta E_{b,form} = I_b * (rand - bias) \qquad (4)$$

Here, $I_b$ is the maximum elastic interaction with zero $bias$, and $rand$ is a random number between 0 and 1. The variable $bias$ determines whether on the average the interaction is positive ($0 < bias < 0.5$) or negative ($0.5 < bias < 1$). In our simulation, we choose $bias = 0.1$ which represents a system that is strongly biased to positive elastic interaction values, as reported in Ref. [11]. Elastic interaction is considered only between the nearest neighbors and since we assume a cubic lattice of bonds at the interface, there are 8 neighboring sites for each siloxane bond. Combining Eq. (3) and Eq. (4), we obtain the final expression for the energy barrier to bond formation

$$E_{b,form}(x,y,t) = E_{b,form,i}(x,y) - \Delta V P(x,y) + \sum_{n=1}^{8} \gamma_n(t) \Delta E_{b,form,n} \qquad (5)$$

where $\Delta E_{b,form,n}$ is the change in energy barrier to bond formation due to the interaction with the $n^{th}$ nearest neighbor site. $\gamma_n = 1$ if bond at $n^{th}$ neighbor reaction site already exists, and otherwise $\gamma_n = 0$. At the beginning of the aging process ($t = 0$), $\gamma_n = 0$ for all reaction sites.

In our model we include the possibility of bond breaking during the stationary contact. Siloxane bond itself is usually considered to be very stable, but strained siloxane bonds (e.g., due to normal pressure in contact) can react with water molecules by hydrolysis reaction [30–32]. During the chemical aging, water molecules can either come from the humid environment [9,13], or be produced during formation of interfacial siloxane bonds. DFT calculations have shown that the range of energy barriers to break an interfacial bond is much narrower than the range of energy barriers to bond formation [see Fig. S3 (b) in Ref. [11]]. In the current model, the energy barrier to bond breaking $E_{b,break}$ is assumed to be a narrow Gaussian distribution centered at 1.1 eV [33], unless noted otherwise.

In our model we also explicitly consider the thermal effects on bond rupture [34]. Specifically, the mean rupture force of a single bond at a constant pulling rate can be expressed as [25]

$$\langle F \rangle = \frac{3E_{b,break}}{2x^{\ddagger}} \left\{ 1 - \left[ \frac{k_B T}{E_{b,break}} \ln \frac{w_0 e^{\gamma} k_B T N(t)}{x^{\ddagger} K_{eff} V} \right]^{2/3} \right\} \tag{6}$$

where $\langle F \rangle$ is the mean bond rupture force, $x^{\ddagger}$ is the distance along the reaction coordinate from a minimum energy to the transition state, $k_B$ is the Boltzmann constant, $T$ is the temperature, $N(t)$ is the number of bonds formed at a hold time $t$, $V$ is the pulling velocity, $\gamma = 0.577$ is the Euler-Mascheroni constant, and $w_0$ is the attempt frequency. $K_{eff} = \frac{N(t)\kappa K}{N(t)\kappa + K}$ is the effective spring constant of the interface, which depends on lateral stiffness of the pulling spring, $K$, and the interfacial stiffness $N(t)\kappa$, where $\kappa$ is the stiffness of a single siloxane bond. In this model, $K_{eff} \approx K$ because the lateral force is applied only after some period of hold time during which the number of formed bonds $N(t)$ increases significantly ($N \approx 40$ at hold time $t = 0.1$ s under normal load $F_N = 23$ nN), resulting in $N(t)\kappa \gg K$.

In experiments, aging $\Delta F$ is measured as the difference between the static and the steady-state kinetic friction force after a certain hold time $t$. Here, we assume that $\Delta F$ comes purely from interfacial bonding during stationary contacts (see Supplemental Material [35] for detailed justification) and therefore $\Delta F = N(t)\langle F \rangle$ [36]. We also ignore reduction in the bond rupture force due to the elastic interaction between neighboring bonds [37], because this effect was found to be negligible in our kMC simulations [35].

We first validated our model by fitting to the recently reported AFM experimental results on load-dependence of aging [13] with different functional forms of $E_{b,form,i}$. We have two fitting parameters: the range and the lower-bound value of the distribution of $E_{b,form,i}$. The results are shown in Fig. 1. The functional form of the distribution shown in Fig. 1 (a) was determined using MD simulations in Ref. [11]. The distribution shown in Fig. 1 (c) assumes that the energy barrier to bond formation increases linearly with the lateral distance between silanol groups from contacting surfaces, which are also explained in more detail in Ref. [11] and its Supplemental Material. We also tried Gaussian and uniform distributions (see Figs. 1 (e) and (g)) and we found that agreement with experiments can be reached irrespectively of the underlying distribution. In general parameters in the model are either set to be consistent with the experiments [13] or chosen within physically reasonable values (see Table. S1 in Supplemental Material [35]). For each type of $E_{b,form,i}$, the simulation results are fitted to the experimental results with only one set of parameters. As a specific example, for distributions obtained from MD [Fig. 1 (a)], we obtain the range of $E_{b,form,i} = 0.68$-$1.3$ eV, and then using the Eyring's relation [Eq. (3)] we obtain the distribution of $E_{b,form}$ (~0.6-1.3 eV). Despite the difference in the forms of those distributions, they all lie in the range of ~0.6-1.4 eV, which is a physically justifiable range [11]. Fig. 1 [(b), (d), (f), and (h)] shows the fitted simulation results of the friction drop $\Delta F$ as a function of the normal load $F_N$ at different hold times along with the experimental data. As it can be seen, the simulation results show linear trends at low loads ($F_N < 400$ nN) no matter what kind of distribution of $E_{b,form,i}$, which all agree well with the experimental data.

At higher normal loads, however, we find deviations from the linear $\Delta F$ vs. $F_N$ relationship. Our simulation results can be fitted with a sublinear function $\Delta F = c_1(F_N + F_{adh})^{2/3} + c_2$ within the entire range of normal loads [Fig. 1 [(b), (d), (f), and (h)], where $c_1$ and $c_2$ are fitting parameters. This is because chemical aging is strongly dependent on the contact area, i.e., a larger contact area provides a larger number of reaction sites available for siloxane bond formation [13]. For adhesive single asperity contact considered here, the area-load relationship is described by the intrinsically non-linear DMT model, whose contact area $A$ increases sublinearly (with the power of 2/3) with the normal load $F_N$ [Eq. (1)]. The fact that friction drop $\Delta F$ shows a non-linear dependence of normal loads in the high load regime fundamentally results from the linear relationship between friction and real contact area $A$.

The non-zero values of $c_2$ obtained from the fits (see Table. S2 in Supplemental Material [35]) indicate that there should be other factors that contribute to $\Delta F$ being not strictly proportional to $A$, i.e., $\Delta F$ is linear with $A$ but it has a small offset in the limit of zero contact area. In our model, these factors can be (1) the pressure dependent energy barrier to bond formation, and (2) the bond rupture force which depends on the number of formed bonds. These effects are discussed next. Nominal (average) normal pressure $\bar{P}$ increases with increasing normal loads due to the sublinear relationship between $A$ and $F_N$. A higher average pressure raises the local contact pressures $P(x, y)$. Based on the Eyring relationship [Eq. (3)], a higher local pressure further reduces the local energy barrier to bond formation, which in turn leads to faster aging. This means that changing the contact area will alter the rates of interfacial chemical reactions and therefore friction, leading to deviations from a linear $\Delta F$–$A$ relationship. In addition, as predicted by Eq. (6), the mean bond rupture force $\langle F \rangle$ decreases with an increasing number of formed bonds $N(t)$. This means that $\langle F \rangle$ is smaller for a larger contact area due to the larger value of $N(t)$, which will also result in the deviations from a linear $\Delta F$–$A$ relationship.

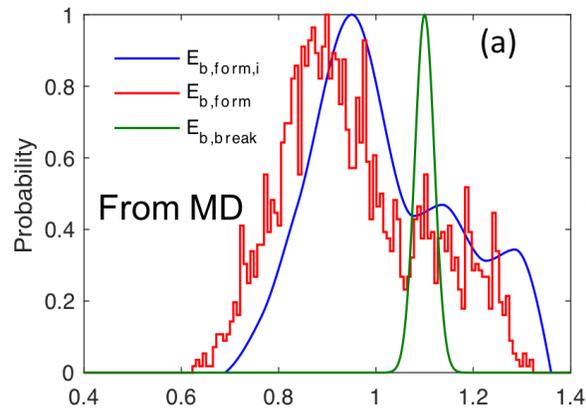
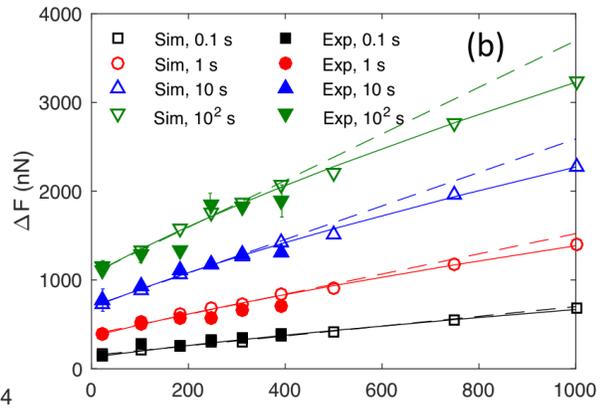
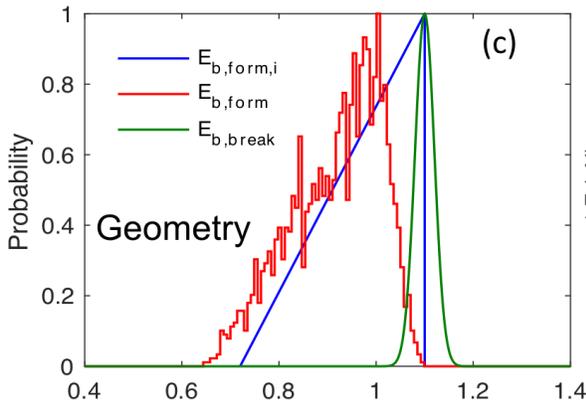
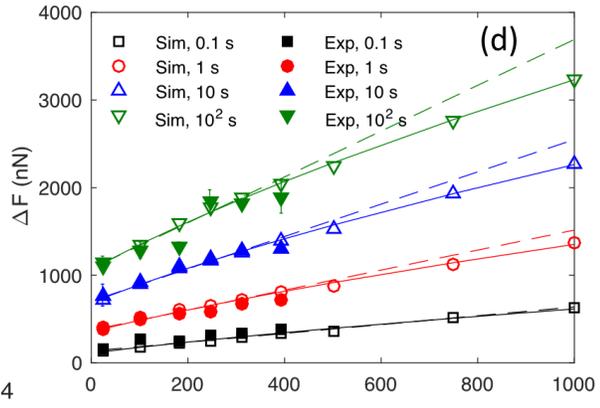
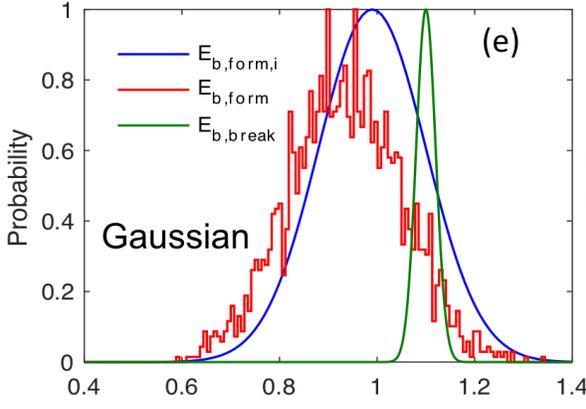
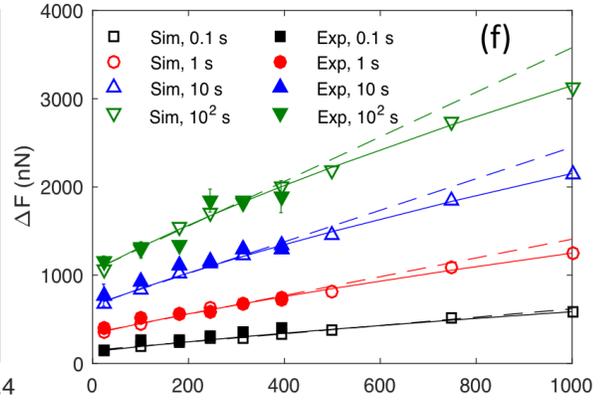
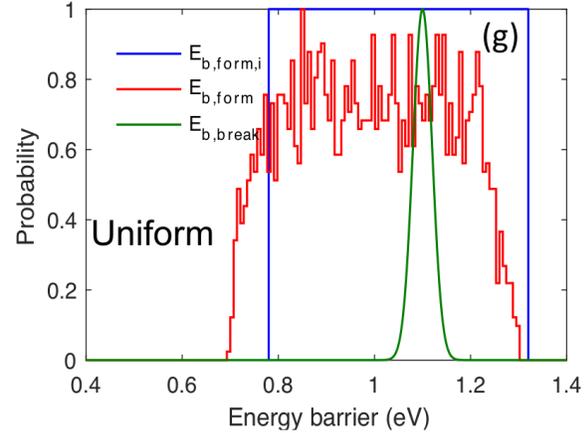
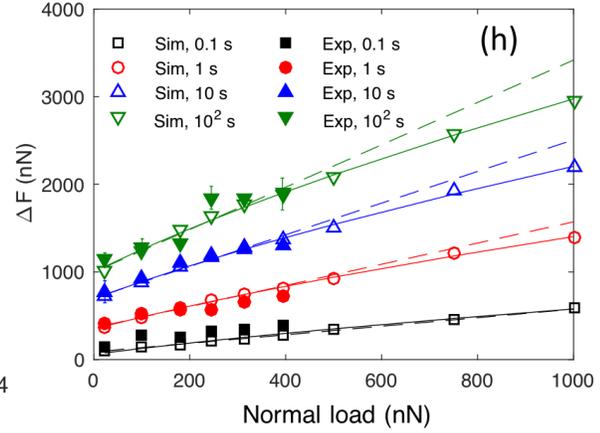

*FIG. 1 Load-dependence of chemical aging with [(a), (b)] distribution obtained from MD, [(c), (d)] geometry distribution (see Ref. [11] and its supplemental materials for details), [(e), (f)] Gaussian distribution, and [(g), (h)]uniform distribution of intrinsic energy barrier to bond formation $E_{b,form,i}$. [(a), (c), (e) and (g)] Initial distribution (t = 0) of energy barriers: blue – intrinsic energy barrier $E_{b,form,i}$ , red – energy barrier to bond formation $E_{b,form}$ under normal load $F_N$ = 393 nN, and green – energy barrier to bond breaking $E_{b,break}$. All distributions are normalized so that the maximum is 1 for easy comparison. [(b), (d), (f), and (h)] Friction drop ΔF as a function of normal load $F_N$ at different hold times t. Hollow symbols and dashed lines are simulation results and linear fits to the simulation results, respectively. Solid lines are power law fits to the simulation results with a function $\Delta F = c_1(F_N+F_{adh})^{2/3}+ c_2$, where $c_1$ and $c_2$ are fitting parameters. Solid symbols correspond to AFM experiments [13].*

Our model allows us to examine the dependence of aging on temperature. The following results are obtained using the same energy barrier distributions as shown in Fig. 1(a). Fig. 2 (a) shows the time-dependence of friction drop ΔF at different temperatures with pulling velocity V = 500 nm/s. We find that the aging process is accelerated as the temperature increases, shifting the logarithmic time dependence regime to shorter time scales. At the same time, the slope $d(\Delta F)/d(\log(t))$ within logarithmic regime decreases and the friction drop ΔF saturates at a lower value as the temperature increases. These results originate from the fact that both chemical bonding during stationary contacts and bond rupturing at the onset of sliding are thermally activated processes. As the temperature increases, more bonds are able to form within a shorter time period, which leads to the acceleration of the aging process. At the same time, however, the bond rupture force ⟨F⟩ decreases due to the thermally assisted escape from the bound state, resulting in the reduction of ΔF. Competition between these two effects results in a non-monotonic temperature dependence of friction drop ΔF, as shown in Fig. 2 (b). The peak in ΔF is at near room temperature, and peak temperature decreases as aging continues (hold time t increases). Similar results can be obtained with a higher pulling velocity, but the peak temperature shifts to slightly higher temperature as shown in Figs. 2 (c) and 2 (d), which indicates that the bond rupture force increases more rapidly with pulling velocity at a higher temperature.

So far, there have been no experimental results reported on temperature dependence of chemical aging in single asperity contacts that can be directly compared to predictions from our model. However, similar AFM experiments were reported in the context of kinetic stick-slip friction where unspecified microscopic bonds are believed to be formed during the stick phase [19]. Much lower peak temperatures (i.e., cryogenic temperatures) were observed in those stick-slip experiments [19], where oxidized silicon tip and substrate are also used, indicating that the microscopic contacts in those experiments may not be siloxane bonds nor other covalent bonds as assumed in our model for stationary contact aging. Indeed, in order to reproduce those experimental results, relatively low energy barriers to the contacts formation ($E_{b,form}$=0.05 eV) and breaking ($E_{b,break}$=0.15 eV) are assumed in the mechano-kinetic model in Ref. [19], which are both too small for siloxane bonds [11,13,33]. The high peak temperature obtained here implies the slow kinetics of covalent bond formation during stationary contacts, which plays an important role especially for long hold times.

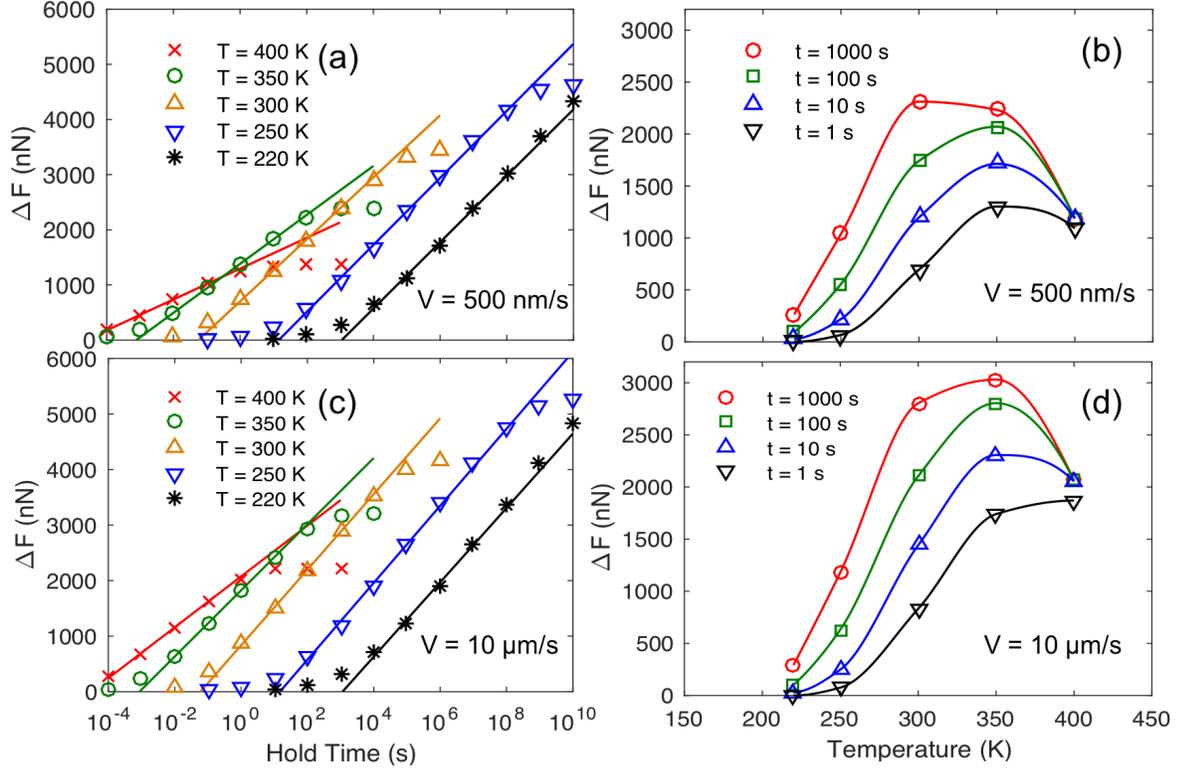

FIG. 2 Temperature effects on chemical aging at $F_N$ = 300 nN. [(a) and (c)] Friction drop $\Delta F$ as a function of hold time t at different temperatures T. [(b) and (d)] $\Delta F$ as a function of temperature T at different hold times t. [(a) and (b)] V = 500 nm/s. [(c) and (d)] V = 10 μm/s.

Our model can also provide understanding of the effects of elastic interaction between neighboring sites and of the energy barrier to bond breaking on frictional aging. Fig. 3 (a) shows how aging $\Delta F$ depends on time for different strengths $I_b$ of elastic interactions. As it can be seen, $\Delta F$ decreases with increasing $I_b$ until it saturates for $I_b$ > 0.5 eV. The monotonic trend for small values of $I_b$ can be easily understood because increasing $I_b$ makes bond formation at neighboring reaction sites more difficult due to the increase in the energy barrier to bond formation $E_{b,form}$, which in turn results in the decrease in total number of formed bonds $N$. When $I_b$ is larger than a certain value, however, most of the bonds can only form at the reaction sites without neighboring bonds, and there will be no more decrease in $N$ (and thus $\Delta F$) with further increase in $I_b$. Interestingly, the increase in the logarithmic of aging on time due to the elastic interaction, which was found in Ref. [11], is not observed in our simulations. One major difference between this model and the original model reported in Ref. [11] is that here we are additionally allowing bonds to break in the stationary contact. In order to examine the effects of bond breaking, we further run simulations with different average values of $E_{b,break}$ as shown in Fig. 3(b). As it can be seen, $E_{b,break}$ indeed affects the time scale of the logarithmic regime, i.e., lower $E_{b,break}$ leads to a shorter logarithmic regime. This is because low $E_{b,break}$ leads to an easier breaking of interfacial bonds, and only reaction sites with $E_{b,form}$ < $E_{b,break}$ have enough time to form interfacial bonds. It is also clear that there is an overall reduction in the value of $\Delta F$ as $E_{b,break}$ decreases. This is due to the reduction in the bond rupture force [Eq. (6)] with decreasing $E_{b,break}$.

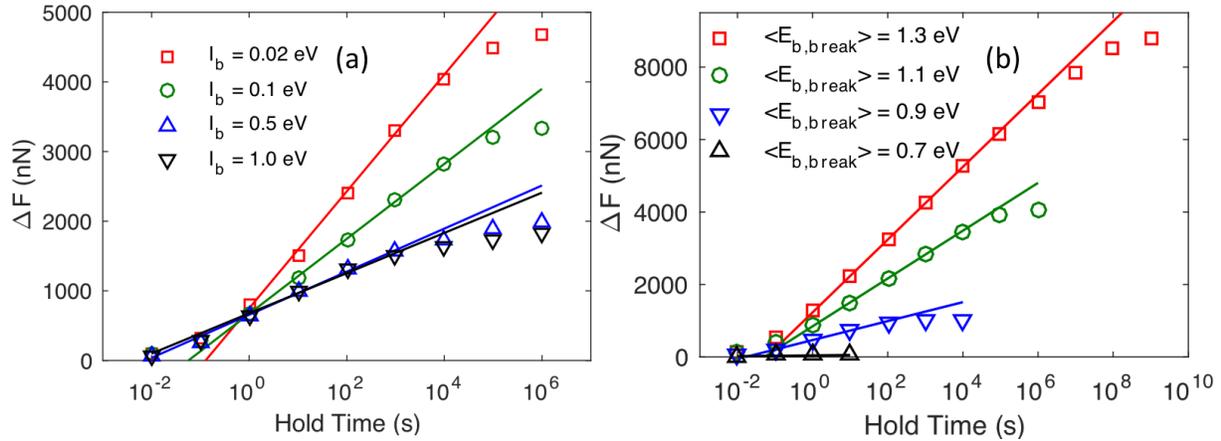

FIG. 3 Time dependence of chemical aging. (a) Friction drop ΔF as a function of hold time with different extent of elastic interaction $I_b$. Note that in Figs. 1 and 2, $I_b = 0.1$ eV is used. (b) Friction drop ΔF as a function of hold time with different mean values of energy barrier to bond breaking $\langle E_{b,break} \rangle$ under $F_N = 393$ nN. In the simulations we keep the standard deviation of the distribution $\langle E_{b,break} \rangle$ constant and it is the same as in Fig. 1 (a). Here, Pulling velocity $V = 500$ nm/s.

In summary, our multi-physics model of chemical aging predicted new phenomena related to load- and temperature-dependence of friction and revealed contributions to aging from specific properties of interfaces. The model shows a quantitative agreement with available AFM experiments on silica-silica interfaces. Our findings should be also applicable to similar materials systems (e.g., other oxides), that form interfacial covalent bonds during stationary contacts and the model can be easily adapted to such materials to determine time-dependent interfacial mechanics.

**ACKNOWLEDGMENT**
The authors gratefully acknowledge financial support from National Science Foundation, grant #1549153 and fruitful discussions with R. Carpick, K. Tian, D. Goldsy and C. Thom from University of Pennsylvania, and Yun Liu from Apple Inc.